# Integrating fuzzy trajectory data and artificial intelligence methods for multi-style lane-changing behavior prediction

**Ruifeng Gu, Research Assistant**

School of Traffic and Transportation Engineering, Central South University,
Changsha, Hunan 410075, P. R. China
Email: guruifeng@Knights.ucf.edu

# Abstract

Artificial intelligence algorithms have been extensively applied in the field of intelligent transportation, especially for driving behavior analysis and prediction. This study proposes a novel framework by integrating fuzzy trajectory data, unsupervised learning and supervised learning methods to predict lane-changing behaviors taking multi driving styles into account. The microscopic trajectory data from the Highway Drone Dataset (HighD) are employed to construct two types of datasets including precise trajectory datasets and fuzzy trajectory datasets for lane-changing prediction models. The fuzzy trajectory data are developed based on different driving styles, which are clustered by the K-means algorithm. Two typical supervised learning methods, including random forest and long-short-term memory combined with convolutional neural network, are further applied for lane-changing behavior prediction. Results indicate that (1) the proposed integration approach performs better than the conventional lane-change prediction; (2) the relative speed-related features have a greater contribution to the lane-changing prediction after being processed by fuzzy rules based on driving styles; and (3) the difference among driving styles is more reflected from the state of lateral movement rather than the lane-changing duration.

# 1. Introduction

Lane-change is a fundamental and complex driving behavior, which occurs when drivers intend to pursue benefits such as faster speeds and comfortable driving conditions on an adjacent lane or drive along a preset path. In recent years, lane-change has attracted increasing attention among researchers due to the high association with the breakdown of traffic flow (Ahn et al., 2010) and various types of traffic crashes (Pande et al., 2006). For example, the number of motor vehicle crashes due to sideswipe (related to lane-change) is approximately 863,100 in the United States in 2018 (NHTSA, 2019). As human error is involved in a large proportion of these lane-change related crashes (NHTSA, 2019), many studies have been devoted to reduce drivers' decision errors such as accurately predicting lane-changing behavior (Xing et al., 2019).

There are three consecutive phases in a typical lane-changing process, which are the formation of intention, maneuver preparation, and performing the maneuver (Leonhardt & Wanielik, 2017). During the lane-keeping status, drivers evaluate the utilities of different lanes and form the lane-changing intention, and then check the surrounding environment to ensure safety during the maneuver preparation phase. Finally, drivers start to change the lane and make adjustments after completing lane-changes. If the lane-change behavior could be accurately predicted before performing the maneuver, more countermeasures and driver assistant systems may be applied to improve driving safety proactively. Therefore, how to develop good lane-changing prediction models to judge lane-change decision becomes worthy of investigations (Song & Li, 2021).

A variety of types of data can be used for lane-changing prediction such as driver's physiological signals, vehicle dynamic data and field driving data. With the advancement of video surveillance and image recognition technologies, an alternative method makes field driving data more accurate and informative, which utilizes microscopic vehicle trajectory data extracted from traffic video footage. Since these videos are usually collected by drones, they provide a bird's-eye view and relatively precise trajectory data of vehicles, which have been extensively utilized in the field of lane-changing prediction (Deo et al., 2018; Chen et al., 2019). Nevertheless, a driver's view is actually different from a bird's-eye view, and the value of features perceived by drivers, like relative speed and distance, is fuzzy instead of precise. What is the difference between fuzzy (a driver's view) and precise (a bird's-eye view) data on lane-changing behavior prediction, and which one performs better for prediction? These questions have not been answered yet and are worth to be explored.

In order to answer the above questions, this study develops a novel framework by integrating fuzzy trajectory data, unsupervised learning and supervised learning methods to predict lane-changing behaviors taking multi-driving styles into account. The microscopic trajectory data from the Highway Drone Dataset (HighD) dataset are employed to construct two types of datasets, including precise trajectory datasets and fuzzy trajectory datasets. Then, a K-means algorithm is applied to classify the lane-changing behaviors into three styles and the fuzzy method of trajectory data is developed based on different driving styles. Two typical supervised learning methods, including random forest (RF) and long-short-term memory combined with convolutional neural network (CNN-LSTM), are further utilized for comparing the lane-changing prediction performance. The current research contributes to existing studies from the following two aspects:

(1) Propose a novel research framework integrating fuzzy trajectory data, unsupervised learning and supervised learning methods to predict lane-changing behaviors;

(2) Compare performance differences between fuzzy (a driver's view) and precise (a bird's-eye view) data on lane-changing behavior prediction taking multi-driving styles into account.

The remainder of this study is organized as follows. The literature review is presented in Section 2. Section 3 describes the dataset and methodology, followed by the results and discussion in Section 4. Section 5 concludes the major findings and provides future research directions.

## 2. Literature review
**2.1 Supervised learning for lane-changing behavior prediction**

Lane-changing behavior prediction has been widely investigated as a binary decision issue which outputs a sign telling whether a vehicle will perform a lane-change (Song & Li, 2021). In recent years, supervised learning approaches (such as support vector machine, decision tree, and random forest) are becoming one of the most appropriate tools for lane-changing prediction due to their good classification performances. Specifically, Dou et al. (2016) used support vector machine (SVM) to predict merge behavior by using NGSIM trajectory data, and the method realized 91% prediction accuracy for non-merge behavior and 84% accuracy for merge behavior. Hu et al. (Hu et al, 2017) presented a decision tree (DT) based method for lane-changing maneuver prediction in cut-in scenarios. Considering a large number of features used in modeling and noises and outliers in datasets, random forest (RF) approach was further applied to classify the driver intention and detect the lane-changing behavior (Schlechtriemen et al., 2015; Deng et al., 2019). With extensive data provided by some microscopic trajectory datasets like Next Generation Simulation (NGSIM) (http://www.ngsim.fhwa.dot.gov) and The Highway Drone Dataset (HighD) (http://www.highhd-dataset.com), deep learning methods also have made tremendous achievements in lane-changing behavior prediction. Since vehicle trajectory data are time-series, many studies utilize recurrent neural network (RNN), which is suitable for time-series problems by containing feedback connections, to predict lane-changing behavior (Xing et al., 2019). One shortage of RNN, however, is that input decays or increases exponentially over time and causes problems in training (Song et al., 2021), so the long short-term memory (LSTM) algorithm is further applied to increase the long-term dependency property and overcome the gradient descent in lane-changing prediction (Guo et al., 2021; Mahajan et al., 2020; Han, et al., 2019).

## 2.2 Input data for lane-changing behavior prediction

There are various types of data for lane-changing behavior prediction and the input data can be divided into three groups according to different resources: traffic context, driver's behavior and physiological signals, and vehicle dynamics (Xing et al., 2019). According to the previous research, lane-changing prediction models can realize high accuracy with the help of driver's behavior and physiological signals data (Yan et al., 2019). However, the data of driver's behavior and physiological signals is difficult to acquire, and some researchers have paid more attention to using the traffic context and vehicle dynamics for lane-changing prediction (Mahajan et al., 2020). Recently, with the advancement of data collection techniques, microscopic trajectory data from video footage have been publicly available. Trajectory datasets have large-scale samples and real-time continuous driving data, including traffic context and vehicle dynamics, and attract more and more researchers to conduct their studies based on these datasets such as NGSIM and HighD (Li et al., 2020; Li et al., 2021; Chen et al., 2021a; Chen et al., 2021b). Trajectory data is high precise and similar to the data detected by seniors mounted on vehicles, so many scholars also utilize trajectory data to develop lane-changing behavior prediction models for advanced driver assistance systems (ADAS) (Deng et al., 2018). For human drivers, however, they predict lane-changing behaviors based on perceived information, which are not as precise as microscopic trajectory data extracted from traffic video software (Perumal et al., 2021). Previously, few attentions have been paid on this issue in the field of trajectory data analysis, so it is inevitable to develop the lane-changing behavior prediction model from drivers' fuzzy view. Although Balal et al. (2016) developed a fuzzy inference system for lane-changing prediction, the linguistic values they used (such as defining the distance as {close, medium, far}) lost much valuable information and the accuracy of lane-changing prediction is not high enough.

According to the above literature review, no study has compared the lane-changing behavior prediction performance between fuzzy and precise trajectory data. When focusing on drivers' fuzzy view, different drivers manifest various driving styles which also need to be taken into account. Therefore, an integrated research framework is inevitable using unsupervised and supervised learning methods for lane-changing behavior prediction based on fuzzy trajectory data.

## 3. Data and methodology

The overall research framework is shown in **Fig. 2**. The framework contains four phases, namely data processing, fuzzy rules setting, input data preparation, and prediction. Section 3.1 introduces the microscopic trajectory data of HighD dataset, and then data processing (Section 3.2) is conducted to extract the lane-changing (LC) decision trajectory dataset from HighD. Driving styles clustering and the fuzzy method are developed in Section 3.3 and 3.4, respectively. The purpose of Section 3.5 is to construct three different datasets as the input data for the classifier. In Section 3.6, two typical supervised learning methods are applied to predict the lane-changing behavior and the performance of different models are evaluated by five performance indicators.

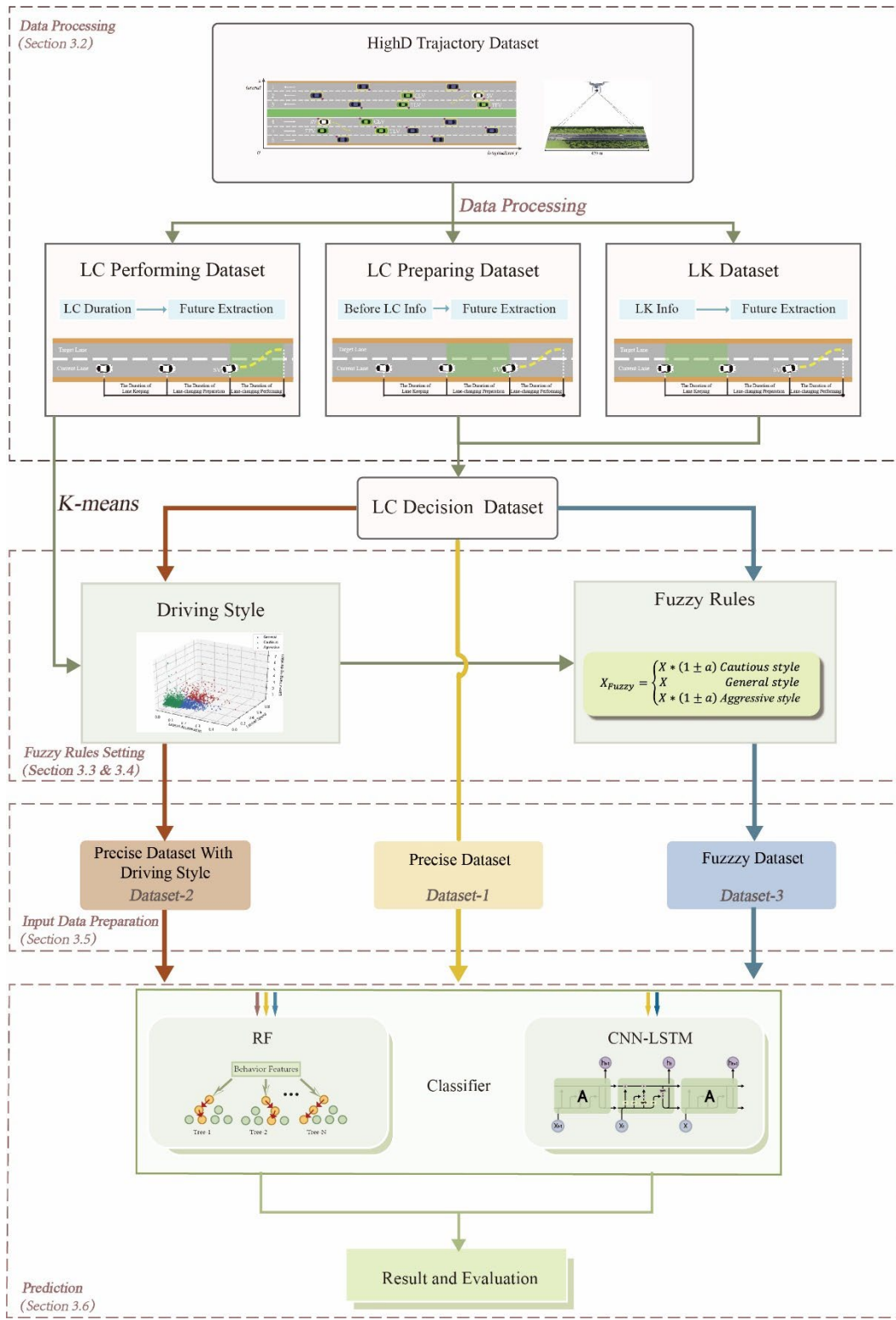

**Fig. 1. The overall framework.**

*Note: LC denotes lane-change.*
1

**3.1 Microscopic vehicle trajectory data**

This study uses the Highway Drone Dataset (HighD) dataset to extract microscopic vehicle trajectory data (Krajewski et al., 2018). The HighD dataset is a naturalistic driving dataset, which is collected via camera-equipped drones from German highways. The collection time of the HighD dataset is on weekdays between 08:00 and 19:00 from 2017 to 2018. There are approximately 16.5 hours of trajectory data with a frame frequency of 25 Hz from six locations near Cologne, Germany. The studied roadways have two or three lanes in each direction.

By using computer vision algorithms and manual annotation methods, the trajectory data are extracted from traffic videos, including frame, lane position, driving direction, lateral and longitudinal velocity, lateral and longitudinal acceleration, information of surrounding vehicles, and others. According to **Fig. 3**, the datapoints of vehicles are placed at the left rear corner of vehicles' bounding boxes. Each lane-changing group involves four vehicles, including the leading vehicle in the target lane (TLV), the following vehicle in the target lane (TFV), and the leading vehicle in the current lane (CLV) as well as the lane-changing subject vehicle (SV).

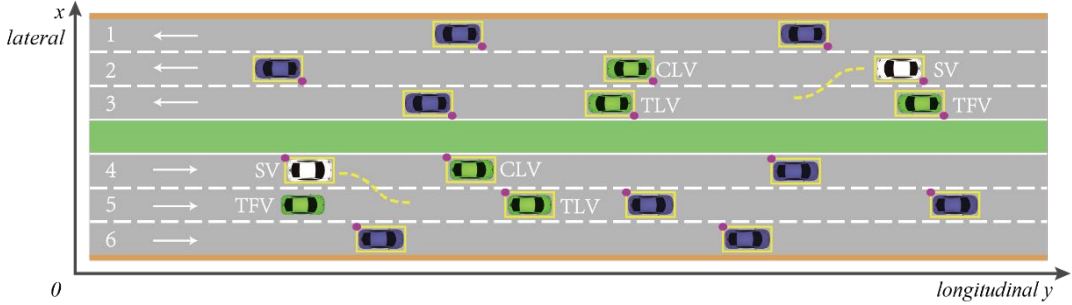

**Fig. 2. Illustration of the trajectory data collection site.**

**3.2 Data processing**

In order to prepare lane-changing decision datasets from raw vehicles' trajectory data, which include the data of lane-changing preparation and lane-keeping process, we first need to pinpoint the time interval of lane-changing operation. The time interval can be calculated via the start time $t_s$ and end time $t_e$ of the operation process. As shown in **Fig. 4**, the HighD dataset provides the position of the left rear point and the width $w$ of the vehicle, and the lane marking line lateral position $D(line)$. By comparing the position of vehicles with the lane line positions, we can obtain the lane-changing start time $t_s$ and the end time $t_e$. In the present study, when the lane number of the SV is changed in the dataset, we define $ID(t)$ as the lane ID used by the lane-changing vehicle at time $t$. When $ID(t) \neq ID(t + \Delta t)$ ($\Delta t$ is the time interval between two consecutive data points), the lane ID changes and we define this moment $t$ as $t_{lc}$.

The second step is to calculate the lateral position difference. We define $D(t)$ as the lateral position difference between the right rear point of the lane-changing vehicle and the left boundaries of the road at time $t$. According to the dataset, the angles between the vehicle and the lane are very small, so the vehicles are assumed to keep parallel to the lane marking lines without considering the angles. Hence, we use $D(t)$ and $D(t) - w$ as the lateral position of the right and left boundaries of the lane-changing vehicle at time $t$, respectively.

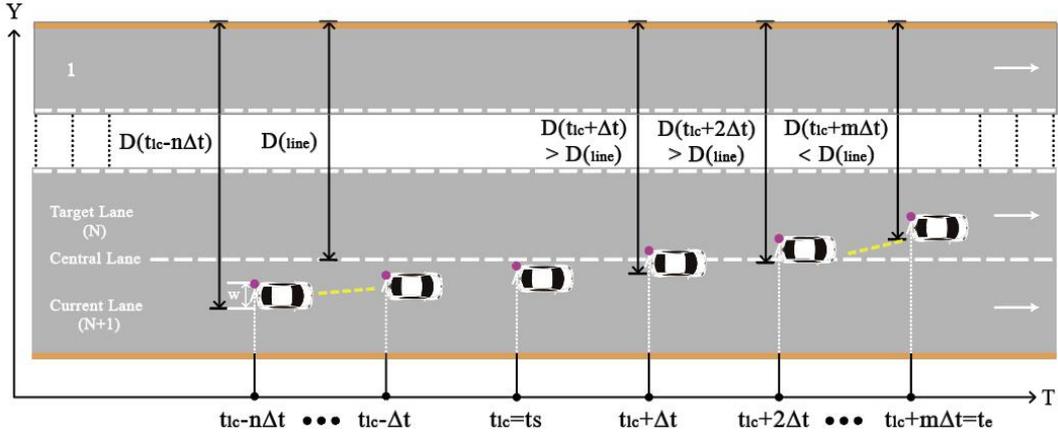

(a) Lane change to the left side

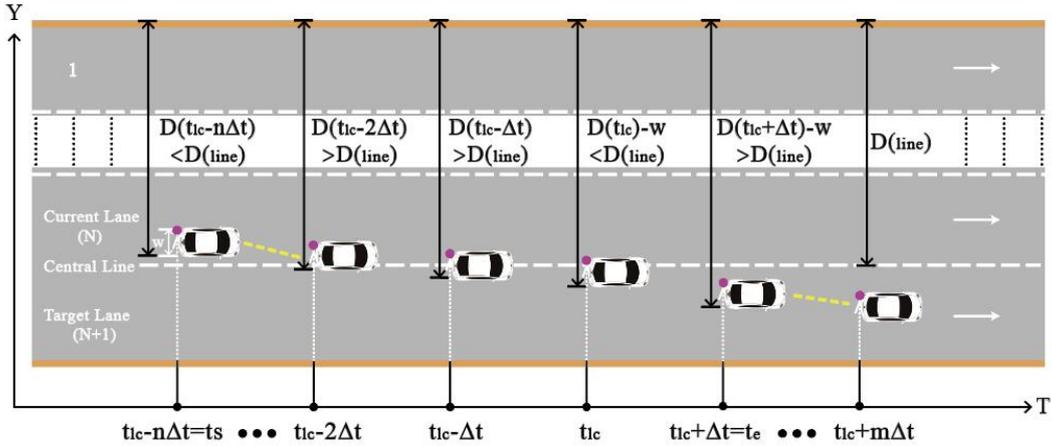

(b) Lane change to the right side

**Fig. 3. Illustration of the lane-changing operation.**

Further, we utilize the lateral position difference to calculate the lane-changing duration. When the vehicle executes left lane-change, the lane-changing start time $t_s$ is $t_{lc}$. After $t_{lc}$, $D(t)$ will still be larger than $D(line)$ for a while and then be smaller. The $t_e$ is the minimum $t$ when $D(t)$ is smaller than $D(line)$, which means the right boundary of the lane-changing vehicle has already passed the central line and driving in the target lane. The method for extracting vehicles changing to the right lane is similar.

It should be noted that the lane-changing duration extracted here only includes the period that the SV crosses two lanes, which is the most critical period of a lane-changing process. The lane-changing start time $t_s$, the end time $t_e$ and and the duration of lane-change $T_{LC}$ can be calculated as follows:

$$t_s = \begin{cases} t_{lc} & (ID(t_{lc}) > ID(t_{lc} + \Delta t)) \\ max\{t|t < t_{lc}\ and\ D(t) - w < D(line)\} & (ID(t_{lc}) < ID(t_{lc} + \Delta t)) \end{cases} \quad (1)$$

$$t_e = \begin{cases} min\{t|t > t_{lc}\ and\ D(t) - w < D(line)\} & (ID(t_{lc}) > ID(t_{lc} + \Delta t)) \\ t_{lc} + \Delta t & (ID(t_{lc}) < ID(t_{lc} + \Delta t)) \end{cases} \quad (2)$$

$$T_{LC} = t_e - t_s \quad (3)$$

With lane-changing duration $T_{LC}$, the data from $t_s$ to $t_e$ can be extracted as lane-change performing dataset. For LC preparing dataset, according to the review by Xing et al. (Xing et al., 2019), the data collected 2-3 s before the maneuver is enough to present the lane-changing intention. Therefore, as shown in **Fig. 5**, 2 s ahead of the lane-changing start time $t_s$ is considered as the duration of lane-changing preparation ($T_{LCP}$) in this research. And 2,276 qualified samples of lane-changing preparation ($T_{LCP}$) constitute the LC preparing dataset. For LK dataset, 2,276 qualified samples of the lane-keeping process ($T_{LK}$) are also included by extracting the duration from $t_2$ to $t_1$, which is also equal to 2 s. Then, the LC preparing dataset and LK dataset constitute the LC decision dataset. LC decision dataset includes a total of 4,552 qualified samples (2,276 lane-change and 2,276 lane-keep). **Table1** shows the description of features considered in this study.

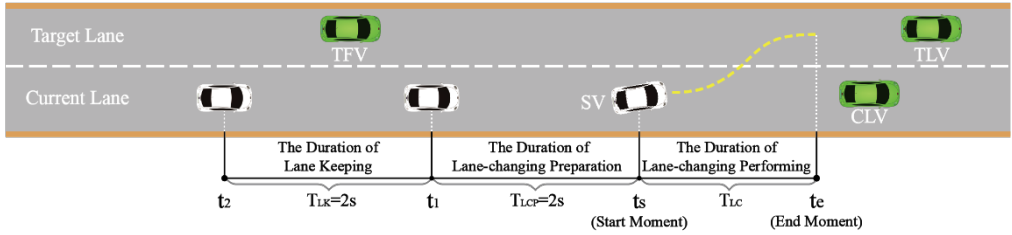

**Fig. 4. Illustration of lane-changing preparation and lane-keeping period**

**Table 1. Variable description.**

| Variable Abbreviation | Variable description |
|---|---|
| ΔY CLV-SV | Vehicle clearance distance between CLV and SV (m) |
| ΔY TLV-SV | Vehicle clearance distance between TLV and SV (m) |
| ΔY SV-TFV | Vehicle clearance distance between SV and TFV (m) |
| ΔV CLV-SV | Vehicle longitudinal speed difference between CLV and SV (m/s) |
| ΔV TLV-SV | Vehicle longitudinal speed difference between TLV and SV (m/s) |
| ΔV SV-TFV | Vehicle longitudinal speed difference between SV and TFV (m/s) |
| $V_Y$ SV | Vehicle longitudinal speed of SV (m/s) |
| $V_Y$ CLV | Vehicle longitudinal speed of CLV (m/s) |
| $V_Y$ TLV | Vehicle longitudinal speed of TLV (m/s) |
| $V_Y$ TFV | Vehicle longitudinal speed of TFV (m/s) |
| $V_X$ SV | Vehicle lateral speed of SV (m/s) |
| $a_Y$ SV | Vehicle longitudinal acceleration of SV (m$^2$/s) |
| $a_Y$ CLV | Vehicle longitudinal acceleration of CLV (m$^2$/s) |
| $a_Y$ TLV | Vehicle longitudinal acceleration of TLV (m$^2$/s) |
| $a_Y$ TFV | Vehicle longitudinal acceleration of TFV (m$^2$/s) |
| $a_X$ SV | Vehicle lateral acceleration of SV (m$^2$/s) |

### 3.3 Driving styles classification based on K-means

Most previous studies identified two to three driving styles based on the degree of aggressiveness or cautiousness (Higgs&Abbas, 2013; Bär et al., 2011; Qi et al., 2015; Ren et al., 2017; de Zepeda et al., 2021). For example, Ren et al. (2017) utilize three different driving styles in the lane-changing model: cautious, stable, and radical. Therefore, we also define three types of diving styles including cautious, general, and aggressive in this study. The K-means method is applied to classify the subject vehicles into three styles based on the selected features of lane-changing behaviors. K-means clustering is one of the most widely used clustering methods (Wang and Xi, 2016). It divides the sample dataset into k subsets denoting k categories, and then assigns samples $D = \{x_1, x_2, x_3, \dots, x_m\}$ into $k$ categories $C = \{C_1, C_2, C_3, \dots, C_k\}$, making each sample has the smallest distance to the category center to which it belongs. The objective function $E$ of k-means clustering can be described as follows:

$$E = \sum_{i=1}^{k} \sum_{x \epsilon C_i} \|x - \mu_i\|^2 \tag{4}$$

$$\mu_i = \frac{1}{|C_i|} \sum_{x \in C_i} x \tag{5}$$

where $\mu_i$ stands for the mean vector of class $C_i$.

### 3.4 Fuzzy method based on different driving styles

To describe the traffic features from drivers' perspective, a fuzzy method based on the different driving styles is developed in this study. We denote that the aggressive drivers consider the lane-changing conditions safe while it is a little risky in the general drivers' view when facing the same lane-changing situations. The cautious drivers may think the lane-change will lead to a crash in the same conditions. Generally, a larger space with lower speed represents a safer environment and a smaller space with faster speed means risky. Therefore, the fuzzy rule is developed as follows:

For distance-related features $X_{distance}$

$$X_{Fuzzy-distance} = \begin{cases} X_{distance} * (1-a) & \text{Cautious style} \\ X_{distance} & \text{General style} \\ X_{distance} * (1+a) & \text{Aggressive style} \end{cases} \quad (6)$$

For speed-related features $X_{speed}$

$$X_{Fuzzy-speed} = \begin{cases} X_{speed} * (1+b) & \text{Cautious style} \\ X_{speed} & \text{General style} \\ X_{speed} * (1-b) & \text{Aggressive style} \end{cases} \quad (7)$$

where $X_{distance}$ and $X_{speed}$ denote the original value of the distance-related and speed-related features extracted from HighD dataset (see Table 1), respectively. The $a$ and $b$ is the fuzzy coefficients whose value are tested from 0.1 to 0.9. It can be found that for cautious style drivers, they perceive the smaller distance-related variables and the larger speed-related variables than the actual ones. The perceived values are opposite for aggressive style drivers. And the optimal fuzzy coefficients' combination is also explored in this research.

**3.5 Input data preparation**

Two types of input data are considered in this research, namely precise data and fuzzy data. Then, three datasets are further prepared as input datasets, including a bird's-eye view dataset (**Dataset-1: precise dataset without driving styles**), a bird's-eye view with driving styles dataset (**Dataset-2: precise dataset considering driving style variable**), and a driver's view dataset (**Dataset-3: fuzzy dataset based on driving styles**). A bird's-eye view dataset (Dataset-1) represents the objective and precise perspective, and the data is identical to LC decision dataset. By labeling the driving style of each vehicle in LC decision dataset, a bird's-eye view dataset with driving style variable (Dataset-2) is produced and it includes not only the precise value dataset (Dataset-1) but also the diving styles of different drivers as an extra features. Fuzzy drivers' view dataset (Dataset-3) is generated from LC decision dataset based on the aforementioned fuzzy method and driving style. The samples used in supervised learning models are split into training (90%) and test (10%) subsets. The standardization and normalization of variables are also implied in the following supervised learning algorithms.

## 3.6 Classifiers and evaluation criteria

The output of lane-changing decision model is a binary value that indicates whether the driver will change the lane, i.e., lane-keep or lane-change. Hence, the lane-changing prediction can be considered as a classification problem. Classifier is a supervised classification learning technique that takes the values of various features of an example and predicts the class label that the example belongs to (Pereira et al., 2009). In this study, two typical supervised learning methods, random forest (RF) and long-short-term memory combined with convolutional neural network (CNN-LSTM), are applied for lane-changing prediction. Note that, we only apply these two methods for illustrating our general research framework and more supervised learning models could also be utilized for the prediction. The framework of proposed lane-changing prediction model is displayed in **Fig. 6**.

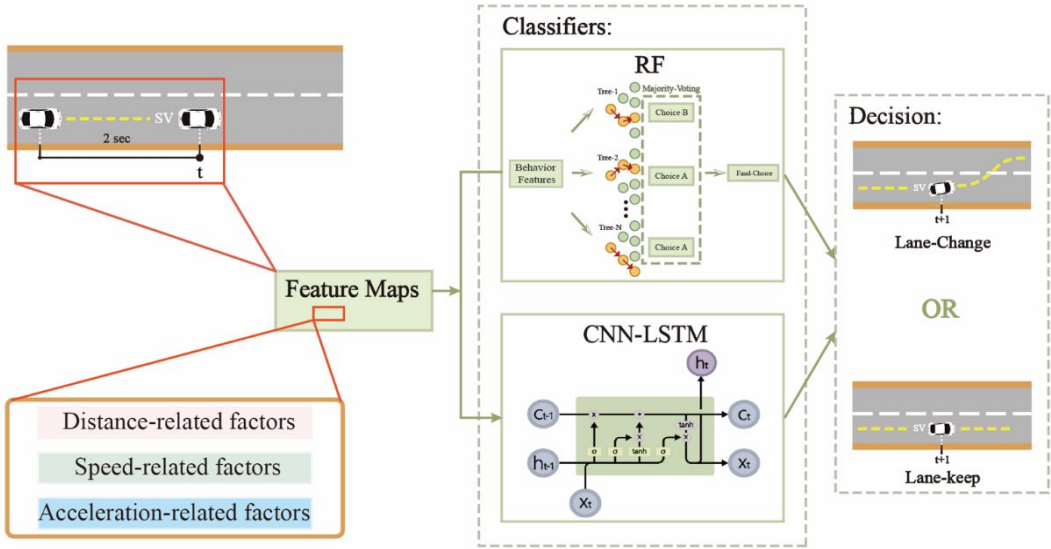

**Fig. 5. The framework of proposed lane-changing prediction model**

As an extension of decision tree, random forest (RF) shows better performance in solving classification problems (Deng et al., 2019). A decision tree poses a series of selection problems, and each final answer to these questions is represented by leaves. Each leaf corresponds to a category in the classification problem. In the present research, RF uses bagging, a bootstrap aggregation technique, to build an ensemble of decision trees as a basic classifier (Breiman, 2001). To construct each tree, the recursive binary splitting method is adopted in which factors are selected to divide the data into different parts. The forecasting result is calculated as follow:

$$P(X) = \frac{1}{N}\sum_{N=1}^{N} Y_N(X) \tag{8}$$

where $Y_k(X)$ is the forecasting result of the kth tree, and k is the number of trees.

LSTM is a kind of recurrent neural network (RNN), which is considered ideal for prediction and classification of time series. Because the LSTM network consists of memory cells so that it can detect dependencies among long-horizon data. LSTM can be trained with a multiple-input single-output model to predict the lane-changing maneuver (Rákos et al., 2020). The input data of LSTM are a window-sized sequence of frames, so the LSTM network is implemented as a classifier to detect lane-changing intentions by classifying two seconds of sequential driving information into lane-change and lane-keep. The calculation process can be constructed as follows:

$$f_t = \sigma(W_{xf}x_t + W_{hf}h_{t-1} + b_f) \tag{9}$$
$$i_t = \sigma(W_{xi}x_t + W_{hi}h_{t-1} + b_i) \tag{10}$$
$$o_t = \sigma(W_{xo}x_t + W_{ho}h_{t-1} + b_o) \tag{11}$$
$$c_t = f_t \odot c_{t-1} + i_i \odot \tanh(W_{xc}x_t + W_{hc}h_{t-1} + b_c) \tag{12}$$
$$h_t = o_t \odot \tanh(c_t) \tag{13}$$

where $f_t$, $i_t$, and $o_t$ are gating vectors, $\sigma(x)$ means sigmoid function, and $\odot$ is the element-wise product.

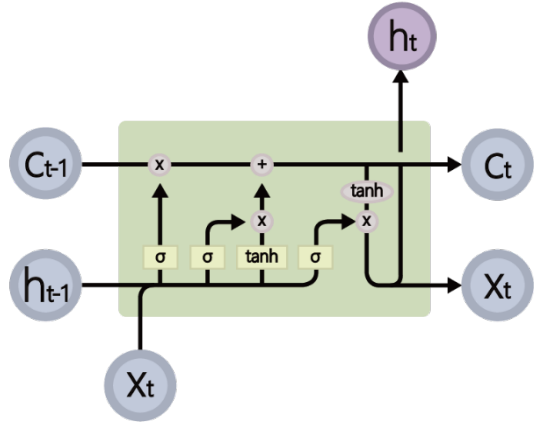

**Fig. 6. The structure of LSTM.**

In this study, CNN layer is applied for feature extraction process of driving information, and integrates with LSTM to constitute CNN-LSTM. The structure of proposed CNN-LSTM is shown in Fig 7, the driving information is sent to CNN layer with ReLU activation function. The pooling layer is added to connect to the CNN layers for reducing the feature size, which contributes to the problem of overfitting during training (Scherer et al., 2010). Furthermore, dropout layers with dropping rate of 0.1 are also inserted to accelerate convergence and mitigate overfitting. Finally, the output of the LSTM is flattened into a dense vector connecting to the dense layer and sigmoid layer for driver lane-changing decision classification.

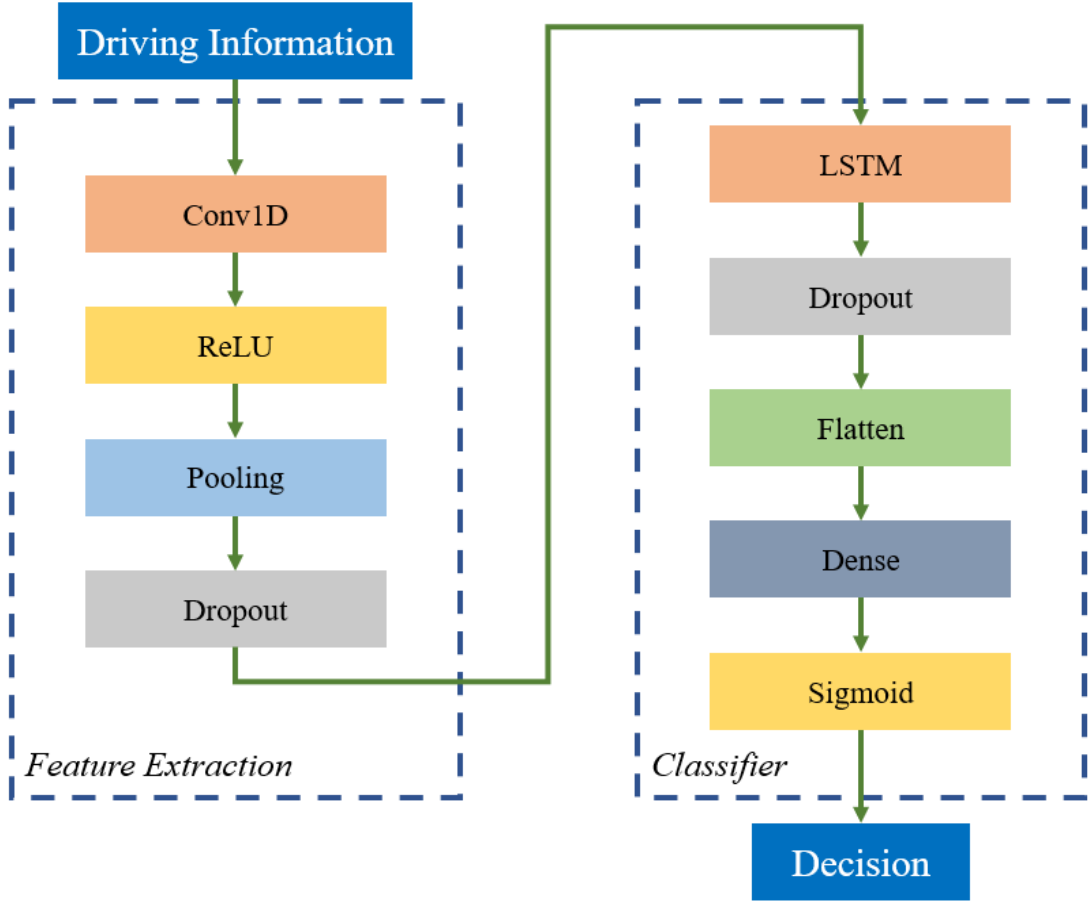

**Fig. 7. Structure of CNN-LSTM.**

The classifiers are evaluated using five performance indicators (including accuracy, precision, recall, F1, and AUC). **Equation (14-17)** shows the formulas of accuracy, precision, recall, and F1, respectively. The AUC can be calculated by the receiver operating characteristic curve (ROC curve).

$$\text{Accuracy} = \frac{TP+TN}{TP+TN+FP+FN} \tag{14}$$

$$\text{Precision} = \frac{TP}{TP+FP} \tag{15}$$

$$\text{Recall} = \frac{TP}{TP+FN} \tag{16}$$

$$F1 = \frac{2*\text{Precision}*\text{Recall}}{\text{Precision}+\text{Recall}} \tag{17}$$

where TP refers to observations correctly identified as lane-change. FN indicates lane-changing samples incorrectly labeled as lane-keep. TN denotes correct prediction of lane-keep, and FP means lane-keep incorrectly predicted as lane-change.

## 4.Result and discussion
### 4.1 Driving style clustering

In this section, K-means method is implemented to classify the lane-changing behaviors of SV into three styles based on the selected features, including lane-changing duration, lateral acceleration of SV, and lateral speed of SV. Results of driving style clustering are shown in **Table 2** and **Fig. 8** presents the distribution of different driving styles in different variables. As shown in **Fig. 8 (a)**, different driving styles have clear boundaries in terms of not only lateral speed but also lateral acceleration. Aggressive drivers have larger lateral speeds and more scattered lateral accelerations, while cautious and general drivers have smaller lateral speeds and the cautious yield smaller lateral accelerations than the general ones. However, no obvious demarcation of driving styles exists in the dimension of lane-changing duration according to **Fig. 8 (b-c)**. The findings demonstrate that the difference among lane-changing driving styles is more reflected from lateral movement (such as sharp speed change) rather than the lane-changing duration.

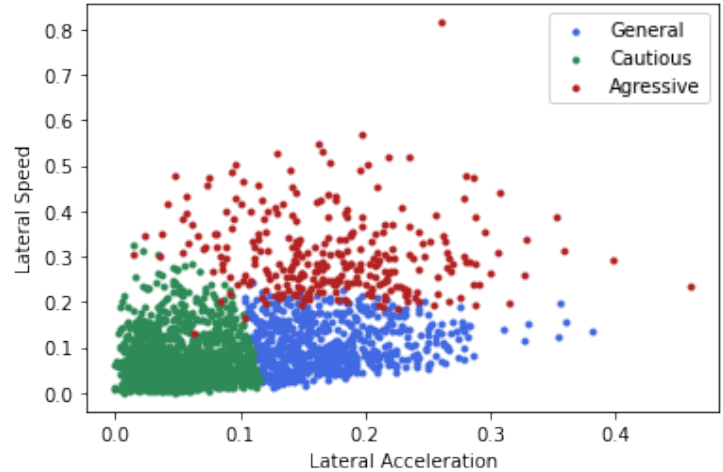

(a)

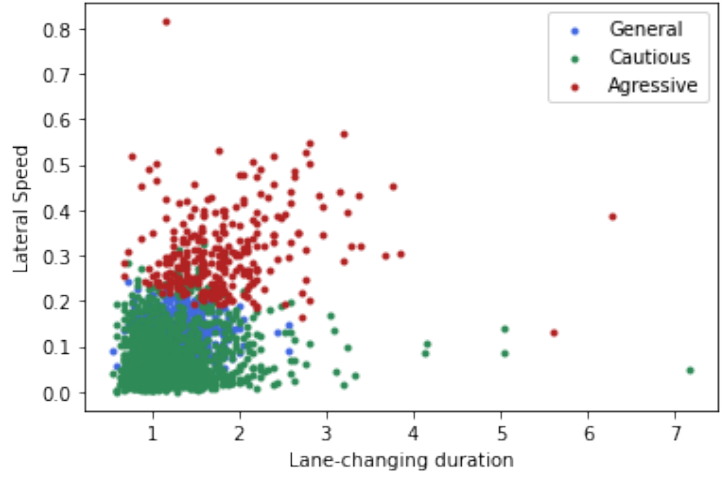

**(b)**

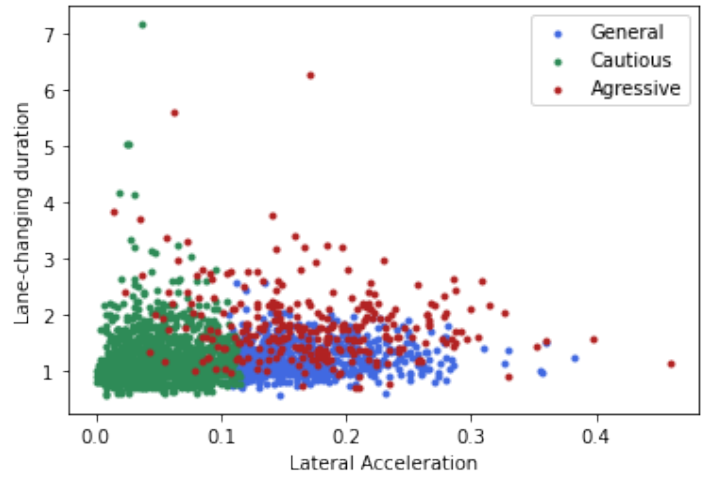

**(c)**

**Fig. 8. The distribution of different driving styles in different variables**

**Table 2 Result of clustering**

| Driving Style | Variables | | | | | |
|---|---|---|---|---|---|---|
| | Duration (s) | | Lateral Acceleration (m) | | Lateral Speed (m) | |
| | Mean | Std. | Mean | Std. | Mean | Std. |
| Aggressive | 1.819 | 0.652 | 0.174 | 0.067 | 0.305 | 0.088 |
| General | 1.235 | 0.289 | 0.052 | 0.029 | 0.072 | 0.056 |
| Cautious | 1.205 | 0.464 | 0.168 | 0.045 | 0.105 | 0.051 |
| Overall | 1.292 | 0.493 | 0.102 | 0.071 | 0.111 | 0.096 |

**4.2 Prediction performance and sensitivity analysis**

The approaches integrating fuzzy trajectory data (Dataset-3) and artificial intelligence methods (RF and CNN-LSTM) with different fuzzy coefficients $a$ and $b$ from 0.1 to 0.9 are developed for lane-changing prediction. The prediction algorithms using a bird's-eye view precise dataset (Dataset-1) represent the conventional lane-change prediction approach which are adopted by many research. The prediction performance of classifiers (RF and CNN-LSTM) and sensitivity analysis of the fuzzy parameters are presented in **Table 3** and **Table 4**. Note that, the 5 groups with the highest accuracy are presented among the 81 kinds of fuzzy coefficients' combinations. Additionally, the input data for RF algorithm is one-dimensional aggregate driving information, including the mean and standard deviation of the selected variables (see **Table 1**). CNN-LSTM algorithm uses sequential driving information data consisting of the value of the variables within two seconds. As the driving style is a categorical variable instead of the sequential driving information, the bird's-eye view dataset considering driving styles (Dataset-2) only used in RF.

**Table 3** displays that the approach which integrates fuzzy trajectory data (Dataset-3) and RF algorithms achieves much better performance in all evaluation criteria than the conventional lane-change prediction using precise trajectory data. Besides, although the approach integrating the bird's-eye view dataset considering driving styles (Dataset-2) does not achieve the highest accuracy, it still performs better than the conventional lane-change prediction without the "help" of driving styles (Dataset-1). The results demonstrate that considering the driving styles is crucial and integrating fuzzy trajectory data with RF is more effective compared with the conventional lane-change prediction using precise trajectory data.

According to the prediction performance of CNN-LSTM (in **Table 4)**, the approach integrating fuzzy trajectory data (Dataset-3) also has better performance than the conventional lane-change prediction with approximately six percentage points improved approximately. It also indicates that the integrated fuzzy method based on different driving styles is feasible both in aggregate feature data and sequential driving data for lane-changing prediction.

We conduct the sensitivity analysis of the different fuzzy coefficients $a$ and $b$ from 0.1 to 0.9. From **Table 3** and **Table 4**, we find that both RF and CNN-LSTM achieve the highest accuracy of 0.943 and 0.965 respectively. Confusion-Matrix and ROC curve of the prediction performance about RF and CNN-LSTM are also presented in **Fig. 9** and **10**. Consequently, the fuzzy coefficients' combinations ($a = 0.6$, $b = 0.1$) and ($a = 0.2$, $b = 0.5$) may be the best fuzzy coefficients for prediction approach using the aggregate feature data and sequential driving data integrating fuzzy method respectively.

Table 3. Prediction performance by RF

| Fuzzy Coefficient | Train set (90%) | | | | | Test set (10%) | | | | |
|---|---|---|---|---|---|---|---|---|---|---|
| | Accuracy | Precision | Recall | F1 | AUC | Accuracy | Precision | Recall | F1 | AUC |
| Bird | 0.992 | 0.992 | 0.992 | 0.992 | 0.943 | 0.882 | 0.886 | 0.882 | 0.882 | 0.943 |
| Bird&DS | 0.996 | 0.996 | 0.996 | 0.996 | 0.960 | 0.893 | 0.894 | 0.893 | 0.893 | 0.960 |
| F_0.6_0.1 | 0.999 | 0.999 | 1.000 | 0.999 | 0.999 | 0.943 | 0.950 | 0.942 | 0.946 | 0.943 |
| F_0.4_0.3 | 0.994 | 0.994 | 0.999 | 0.990 | 0.994 | 0.941 | 0.957 | 0.929 | 0.943 | 0.941 |
| F_0.2_0.4 | 0.999 | 0.999 | 1.000 | 0.999 | 0.999 | 0.939 | 0.938 | 0.946 | 0.942 | 0.938 |
| F_0.5_0.4 | 0.998 | 0.998 | 1.000 | 0.996 | 0.998 | 0.939 | 0.957 | 0.925 | 0.941 | 0.939 |
| F_0.6_0.4 | 0.994 | 0.994 | 0.996 | 0.992 | 0.994 | 0.936 | 0.934 | 0.946 | 0.940 | 0.936 |

*Note: Bird and Bird&DS represent the Bird's view dataset (Dataset-1) and Bird's view with driving styles dataset (Dataset-2), respectively. F_a_b represent the Fuzzy drivers' view datasets (Dataset-3) with different fuzzy coefficients $a$ and $b$ (from 0.1 to 0.9).*

Table 4. Prediction performance by CNN-LSTM

| Fuzzy Coefficient | Train set (90%) | | | | | Test set (10%) | | | | |
|---|---|---|---|---|---|---|---|---|---|---|
| | Accuracy | Precision | Recall | F1 | AUC | Accuracy | Precision | Recall | F1 | AUC |
| Bird | 0.892 | 0.902 | 0.878 | 0.890 | 0.892 | 0.899 | 0.920 | 0.881 | 0.900 | 0.890 |
| F_0.2_0.5 | 0.977 | 0.973 | 0.981 | 0.977 | 0.977 | 0.965 | 0.955 | 0.979 | 0.967 | 0.964 |
| F_0.2_0.1 | 0.977 | 0.970 | 0.984 | 0.977 | 0.977 | 0.963 | 0.962 | 0.966 | 0.964 | 0.963 |
| F_0.5_0.1 | 0.977 | 0.982 | 0.972 | 0.977 | 0.977 | 0.963 | 0.966 | 0.962 | 0.964 | 0.963 |
| F_0.4_0.3 | 0.978 | 0.971 | 0.986 | 0.978 | 0.978 | 0.963 | 0.954 | 0.975 | 0.964 | 0.962 |
| F_0.4_0.1 | 0.976 | 0.970 | 0.982 | 0.976 | 0.976 | 0.961 | 0.962 | 0.962 | 0.962 | 0.960 |

*Note: Bird represents the Bird's view dataset (Dataset-1). F_a_b represent the Fuzzy drivers' view datasets (Dataset-3) with different fuzzy coefficients $a$ and $b$ (from 0.1 to 0.9).*

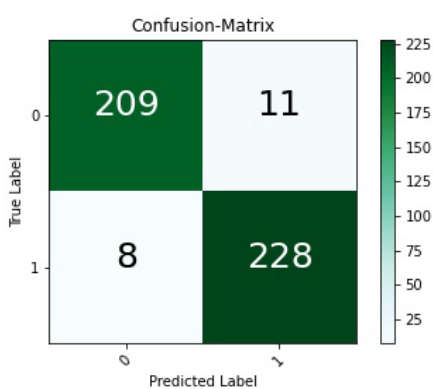
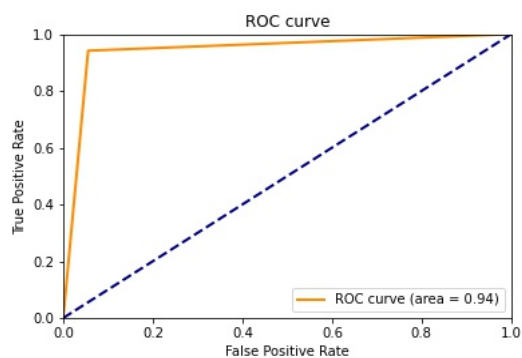

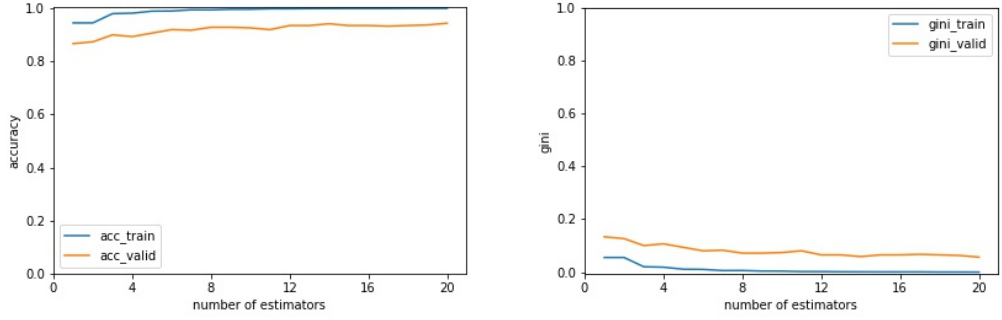

**Fig. 9. Prediction performance and learning curve of RF with fuzzy coefficient (0.6, 0.1) based on the test set.**

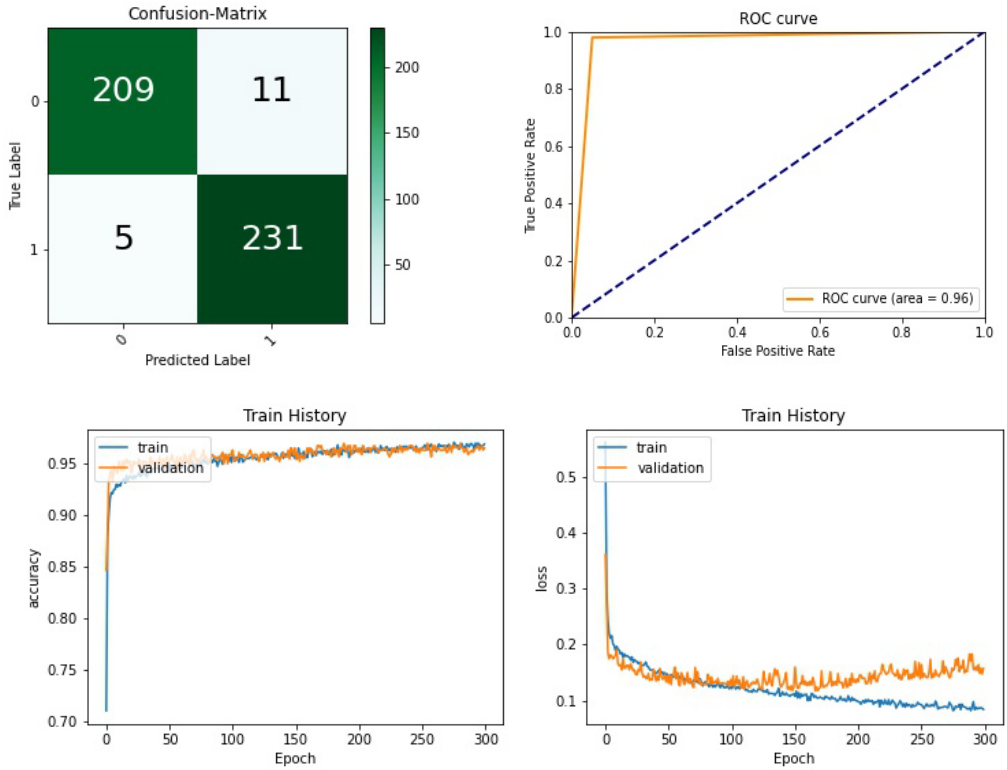

**Fig. 10. Prediction performance and learning curve of CNN-LSTM with fuzzy coefficient (0.2, 0.5) based on the test set.**

### 4.3 Feature importance

To explore how the integrated research method improves the prediction performance, the importance of all input features of the RF method is estimated. **Table 5** shows the top five features in each dataset sorted by features' importance. We can find that the important features of a bird's-eye view dataset (Dataset-1) are similar to a bird's-eye view dataset considering driving styles (Dataset-2) with only one different feature, which proves the necessity of considering driving styles in lane-changing prediction. In the integrated fuzzy method with fuzzy drivers' view dataset (Dataset-3), the standard deviation of ΔV TLV-SV, ΔV CLV-SV, and ΔV SV-TFV show significance in lane-changing prediction compared with the precise value datasets. The result evinces that the relative speed-related features have a greater contribution to the lane-changing prediction from the drivers' view when taking different driving styles into account. The possible reason may be that the influence of relative speed-related features on lane-changing decision varies from different driving styles and relative speed-related variables may have strong individual heterogeneity in lane-changing prediction. Additionally, although the tendency of $V_X$ SV's importance decreases after the trajectory data is fuzzy based on the driving styles, the lateral speed of SV is still the most important feature for detecting the lane-changing intention.

**Table 5. Feature importance**

| Dataset | $V_X$ SV (mean) | ΔV TLV-SV (std.) | ΔV CLV-SV (std.) | ΔV SV-TFV (std.) | $a_X$ SV (mean) | $a_X$ SV (std.) | ΔY SV-TFV (mean) | $V_X$ SV (std.) | Driving Style |
|---|---|---|---|---|---|---|---|---|---|
| **Bird** | 0.478 | - | - | - | 0.062 | 0.047 | 0.031 | 0.026 | / |
| **Bird&DS** | 0.410 | - | - | - | 0.085 | 0.051 | - | 0.043 | 0.045 |
| **F_0.6_0.1** | 0.315 | 0.030 | - | 0.054 | 0.087 | 0.104 | - | - | / |
| **F_0.4_0.3** | 0.287 | | | 0.043 | 0.060 | 0.062 | - | 0.166 | / |
| **F_0.2_0.4** | 0.349 | | 0.055 | 0.035 | 0.076 | - | - | 0.115 | / |
| **F_0.5_0.4** | 0.269 | | 0.031 | 0.064 | 0.072 | - | - | 0.179 | / |
| **F_0.6_0.4** | 0.259 | 0.038 | | 0.067 | 0.102 | - | - | 0.161 | / |

*Note: Bird and Bird&DS represent the Bird's view dataset (Dataset-1) and Bird's view with driving styles dataset (Dataset-2), respectively. F_a_b represent the Fuzzy drivers' view datasets (Dataset-3) with different fuzzy coefficients a and b (from 0.1 to 0.9). Additionally, Driving style is not included in Bird's view dataset (Dataset-1) and Fuzzy drivers' view datasets (Dataset-3).*

# 5. Conclusions

This study proposed a novel framework by integrating fuzzy trajectory data, unsupervised learning and supervised learning methods to predict lane-changing behaviors. We compared the performance of the approach integrating fuzzy trajectory data (Dataset-3) and artificial intelligence methods with the conventional lane-change prediction using precise data (Dataset-1&2). For the artificial intelligence methods, RF and CNN-LSTM algorithms are applied to predict the lane-changing behavior, and the K-means method is applied for driving style clustering. Three datasets including precise datasets (a bird's view dataset and a bird's view with driving styles dataset) and fuzzy datasets (drivers' view dataset) are taken into account to examine their performance on the lane-changing prediction. The major conclusions are as follows:

(1) The proposed approach integrating fuzzy data performs better than the conventional lane-change prediction using precise data. No matter in using aggregate feature data and sequential driving data for lane-changing prediction, the data processed by fuzzy rules show a higher accuracy of lane-changing behavior prediction.

(2) The relative speed-related features show a greater contribution to the lane-changing prediction from the driver's view. It demonstrates that relative speed-related variables are worth being fuzzy in the integrated framework and may have strong individual heterogeneity in lane-changing prediction.

(3) The difference among lane-changing driving styles is more reflected in the state of lateral movement rather than the lane-changing duration. Clear boundaries exist in different driving styles in the dimension of lateral speed and lateral acceleration, while no demarcation of driving styles exists in the dimension of lane-changing duration.

The aforementioned findings reveal the feasibility and advantages of the fuzzy method based on different driving styles in lane-changing behavior prediction and propose a novel framework by integrating fuzzy trajectory data, unsupervised learning and supervised learning methods. The automated vehicles can also utilize these findings as references in their artificial intelligence algorithms to improve the accuracy of the driving intention determination among surrounding manually-driven vehicles. Considering practical application instead of the ideally connected environment, such as the limitation of the sensor, the application of these findings is worthy of being investigated in the future. Besides, owing to the limitation of trajectory datasets, we did not test the performance of fuzzy data in the urban road environment. In the future, this work needs to be further conducted since the urban road environment is more complex and riskier compared to freeways.

**Acknowledgments**

   This research was sponsored by the National Natural Science Foundation of China (71901223, 52102405), and Natural Science Foundation of Hunan Province (2021JJ40746, 2021JJ40603).

# REFERENCES

Ahn, S., Laval, J. (2010). Cassidy, M.J. Effects of merging and diverging on freeway traffic oscillations. Transportation Research Record, (2188), pp. 1-8.

Balal, E., Cheu, R. L., & Sarkodie-Gyan, T. (2016). A binary decision model for discretionary lane changing move based on fuzzy inference system. Transportation Research Part C: Emerging Technologies, 67, 47-61.

Bär, T., Nienhüser, D., Kohlhaas, R., & Zöllner, J. M. (2011, October). Probabilistic driving style determination by means of a situation based analysis of the vehicle data. In 2011 14th International IEEE Conference on Intelligent Transportation Systems (ITSC) (pp. 1698-1703). IEEE.

Breiman, L., 2001. Random forests. Mach. Learn. 45, 5–32.

Chen, Q., Gu, R., Huang, H. , Lee, J. , Zhai, X. , & Li, Y. (2021a). Using vehicular trajectory data to explore risky factors and unobserved heterogeneity during lane-changing. Accident Analysis & Prevention, 151.

Chen, Q., Huang, H., Li, Y., Lee, J., Long, K., Gu, R., & Zhai, X. (2021b). Modeling accident risks in different lane-changing behavioral patterns. *Analytic methods in accident research*, *30*, 100159. Ahn, S., Laval, J. (2010). Cassidy, M.J. Effects of merging and diverging on freeway traffic oscillations. Transportation Research Record, (2188), pp. 1-8.

Chen, Y., Hu, C., & Wang, J. (2019). Motion planning with velocity prediction and composite nonlinear feedback tracking control for lane-change strategy of autonomous vehicles. IEEE Transactions on Intelligent Vehicles, 5(1), 63-74.

de Zepeda, M. V. N., Meng, F., Su, J., Zeng, X. J., & Wang, Q. (2021). Dynamic clustering analysis for driving styles identification. Engineering Applications of Artificial Intelligence, 97, 104096.

Deng, Q., Wang, J., & Soffker, D. (2018, June). Prediction of human driver behaviors based on an improved HMM approach. In 2018 IEEE Intelligent Vehicles Symposium (IV) (pp. 2066-2071). IEEE.

Deng, Q., Wang, J., Hillebrand, K., Benjamin, C. R., & Söffker, D. (2019). Prediction performance of lane changing behaviors: a study of combining environmental and eye-tracking data in a driving simulator. IEEE Transactions on Intelligent Transportation Systems, 21(8), 3561-3570.

Deng, Q., Wang, J., Hillebrand, K., Benjamin, C. R., & Söffker, D. (2019). Prediction performance of lane changing behaviors: a study of combining environmental and eye-tracking data in a driving simulator. IEEE Transactions on Intelligent Transportation Systems, 21(8), 3561-3570.

Deo, N., & Trivedi, M. M. (2018, June). Multi-modal trajectory prediction of surrounding vehicles with maneuver based lstms. In 2018 IEEE Intelligent Vehicles Symposium (IV) (pp. 1179-1184). IEEE..

Dou, Y., Yan, F., & Feng, D. (2016, July). Lane changing prediction at highway lane drops using support vector machine and artificial neural network classifiers. In 2016 IEEE International Conference on Advanced Intelligent Mechatronics

(AIM) (pp. 901-906). IEEE.

Guo, Y., Zhang, H., Wang, C., Sun, Q., & Li, W. (2021). Driver lane change intention recognition in the connected environment. Physica A: Statistical Mechanics and its Applications, 575, 126057.

Han, T., Jing, J., & Özgüner, Ü. (2019, June). Driving intention recognition and lane change prediction on the highway. In 2019 IEEE Intelligent Vehicles Symposium (IV) (pp. 957-962). IEEE.

Higgs, B., & Abbas, M. (2013, October). A two-step segmentation algorithm for behavioral clustering of naturalistic driving styles. In 16th International IEEE Conference on Intelligent Transportation Systems (ITSC 2013) (pp. 857-862). IEEE

Hu, M., Liao, Y., Wang, W., Li, G., Cheng, B., & Chen, F. (2017). Decision tree-based maneuver prediction for driver rear-end risk-avoidance behaviors in cut-in scenarios. Journal of advanced transportation, 2017.

Kala, R., & Warwick, K. (2013). Motion planning of autonomous vehicles in a non-autonomous vehicle environment without speed lanes. Engineering Applications of Artificial Intelligence, 26(5-6), 1588-1601

Kala, R., & Warwick, K. (2013). Motion planning of autonomous vehicles in a non-autonomous vehicle environment without speed lanes. Engineering Applications of Artificial Intelligence, 26(5-6), 1588-1601.

Krajewski, R., Bock, J., Kloeker, L., & Eckstein, L. (2018, November). The highd dataset: A drone dataset of naturalistic vehicle trajectories on german highways for validation of highly automated driving systems. In 2018 21st International Conference on Intelligent Transportation Systems (ITSC) (pp. 2118-2125). IEEE.

Leonhardt, V., & Wanielik, G. (2017, July). Feature evaluation for lane change prediction based on driving situation and driver behavior. In 2017 20th International Conference on Information Fusion (Fusion) (pp. 1-7). IEEE.

Li, Y., Gu R., Lee J., Yang M., Chen Q., Zhang Y. (2021). The dynamic tradeoff between safety and efficiency in discretionary lane-changing behavior: a random parameters logit approach with heterogeneity in means and variances. Accident Analysis and Prevention, 153(2021) 106036.

Li, Y., Wu D., Lee J., Yang M., Shi Y. (2020). Analysis of the transition condition of rear-end collisions using time-to-collision index and vehicle trajectory data. Accident Analysis and Prevention, 144.

Mahajan, V., Katrakazas, C., & Antoniou, C. (2020). Prediction of lane-changing maneuvers with automatic labeling and deep learning. Transportation research record, 2674(7), 336-347.

Mahajan, V., Katrakazas, C., & Antoniou, C. (2020). Prediction of lane-changing maneuvers with automatic labeling and deep learning. Transportation research record, 2674(7), 336-347.

NHTSA. (2019). Traffic Safety Facts Annual Report Tables 2019. U.S. 26 Department of Transportation

Pande, A., & Abdel-Aty, M. (2006). Assessment of freeway traffic parameters leading to lane-change related collisions. Accident Analysis & Prevention, 38(5), 936–948.

Pereira, F., Mitchell, T., Botvinick, M., 2009. Machine learning classifiers and fMRI: a tutorial overview. Neuroimage 45, S199–S209

Perumal, P. S., Sujasree, M., Chavhan, S., Gupta, D., Mukthineni, V., Shimgekar, S. R., ... & Fortino, G. (2021). An insight into crash avoidance and overtaking advice systems for Autonomous Vehicles: A review, challenges and solutions. Engineering applications of artificial intelligence, 104, 104406.

Qi, G., Du, Y., Wu, J., & Xu, M. (2015). Leveraging longitudinal driving behaviour data with data mining techniques for driving style analysis. IET intelligent transport systems, 9(8), 792-801.

Rákos, O., Aradi, S., & Bécsi, T. (2020). Lane Change Prediction Using Gaussian Classification, Support Vector Classification and Neural Network Classifiers. Periodica Polytechnica Transportation Engineering, 48(4), 327-333.

Ren, G., Zhang, Y., Liu, H., Zhang, K., & Hu, Y. (2019). A new lane-changing model with consideration of driving style. International Journal of Intelligent Transportation Systems Research, 17(3), 181-189.

Scherer, D., Müller, A., & Behnke, S. (2010, September). Evaluation of pooling operations in convolutional architectures for object recognition. In International conference on artificial neural networks (pp. 92-101). Springer, Berlin, Heidelberg.

Schlechtriemen, J., Wirthmueller, F., Wedel, A., Breuel, G., & Kuhnert, K. D. (2015, June). When will it change the lane? A probabilistic regression approach for rarely occurring events. In 2015 IEEE Intelligent Vehicles Symposium (IV) (pp. 1373-1379). IEEE.

Song, R., & Li, B. (2021). Surrounding Vehicles' Lane Change Maneuver Prediction and Detection for Intelligent Vehicles: A Comprehensive Review. IEEE Transactions on Intelligent Transportation Systems.

Wang, W., & Xi, J. (2016, July). A rapid pattern-recognition method for driving styles using clustering-based support vector machines. In 2016 American Control Conference (ACC) (pp. 5270-5275). IEEE.

Xing, L., He, J., Abdel-Aty, M., Cai, Q., Li, Y., & Zheng, O., 2019. Examining traffic conflicts of up stream toll plaza area using vehicles' trajectory data. Accident Analysis & Prevention, 125, 174-187.

Xing, Y., Lv, C., Wang, H., Wang, H., Ai, Y., Cao, D., ... & Wang, F. Y. (2019). Driver lane change intention inference for intelligent vehicles: framework, survey, and challenges. IEEE Transactions on Vehicular Technology, 68(5), 4377-4390.

Yan, Z., Yang, K., Wang, Z., Yang, B., Kaizuka, T., & Nakano, K. (2019, June). Time to lane change and completion prediction based on Gated Recurrent Unit Network. In 2019 IEEE Intelligent Vehicles Symposium (IV) (pp. 102-107). IEEE.